\documentclass[12pt,a4paper]{article}
\usepackage{epsfig}
\def\ga{\mathrel{\mathpalette\fun >}}
\def\fun#1#2{\lower3.6pt\vbox{\baselineskip0pt\lineskip.9pt
\ialign{$\mathsurround=0pt#1\hfil##\hfil$\crcr#2\crcr\sim\crcr}}}

\begin{document}
\begin{titlepage}
\thispagestyle{empty}

\hfill ITEP-PH-2/2000 

\hfill IFIC-00/39 

\hfill SINP MSU 2000-20/624

\hfill \today

\vspace*{1cm}   
\begin{center}   
\Large \bf On the search for 50 GeV neutrinos
\end{center}

\vspace*{1cm}  

\begin{center}
\large V.A.~Ilyin \\
{\normalsize \it SINP MSU, 119899 Moscow, Russia} \\
{\normalsize \sf ilyin@theory.npi.msu.su}
\end{center}     

\begin{center}
\large M.~Maltoni \\
{\normalsize \it Instituto de F\'{\i}sica Corpuscular~--~C.S.I.C.,} \\
{\normalsize \it Departament de F\'{\i}sica Te\`orica, 
        Universitat de Val\`encia,} \\
{\normalsize \it Edificio Institutos de Paterna, Apt.~22085,
   E-46071 Valencia, Spain} \\ 
{\normalsize \sf maltoni@ific.uv.es}
\end{center}     

\begin{center}
\large V.A.~Novikov, L.B~Okun \\
{\normalsize \it ITEP, 117218 Moscow, Russia} \\
{\normalsize \sf novikov@heron.itep.ru} \hspace{0.5cm}
{\normalsize \sf okun@heron.itep.ru}
\end{center}     

\begin{center}
\large A.N.~Rozanov \\
{\normalsize \it ITEP and CPPM, IN2P3-CNRS, Univ. M\'{e}diterran\'{e}e, F-13288,
   Marseilles, France} \\
{\normalsize \sf rozanov@cppm.in2p3.fr}
\end{center}     

\begin{center}
\large M.I.~Vysotsky \\
{\normalsize \it ITEP, 117218 Moscow, Russia} \\
{\normalsize \sf vysotsky@heron.itep.ru}
\end{center}     

\vspace{2cm}

\begin{abstract}
Using the computer code CompHEP we estimate the number of events and the
background, at LEP II and TESLA, for the reaction $e^+ e^- \to
N\bar{N}\gamma$, where $N$ is a hypothetical Dirac neutrino with mass of the
order of 50 GeV.
\end{abstract}

\end{titlepage}

\newpage

\vspace{0.5cm}
{\large \bf 1. Introduction}
\vspace{0.3cm}

The precision measurements of ``invisible width" of $Z$-bosons prove that
there exist only three ordinary neutrinos: $\nu_e$, $\nu_{\mu}$,
$\nu_{\tau}$ \cite{1}. Precision measurements of other $Z$-observables when
compared with calculations of radiative corrections exclude the existence
of the fourth generation of fermions in the case when the fourth neutrino
$N$ is heavy ($M_N \ga 100$ GeV) \cite{2}. However, as was shown recently
\cite{3}, the radiative corrections become small if $m_N \sim m_Z/2$. Thus
the existence of fourth generation with charged lepton and quarks being
heavy ($M\ga 100$ GeV) but with neutrino being ``light" ($M_N \sim 50$ GeV)
is still an open possibility. 

As was proposed a long time ago \cite{6}, the cross section of
$e^+e^-$-annihilation into an invisible final state can be inferred from
observation of initial state bremsstrahlung. In LEP II experiments
\cite{ALEPH,DELPHI,L3,OPAL} the production of final states involving one
energetic photon, 
\begin{equation}
e^+ e^- \to \gamma + \; {\rm "nothing"} \;\; ,
\label{1}
\end{equation}
was studied at centre-of-mass energies up to 202 GeV. In context of our
discussion ``nothing" means: a) $\nu_\mu\bar\nu_\mu$ and $\nu_\tau
\bar\nu_\tau$ produced in decays of real and virtual $Z$ bosons, b) $\nu_e
\bar\nu_e$, where two mechanisms contribute, through $s$-channel $Z$ boson
and from $t$-channel exchange of $W$ boson, and c) $N\bar N$ pair from
virtual $Z$ bosons. 

The aim of this note is to present results of calculation of the
differential cross sections of the reaction (\ref{1}) for various values of
$m_N$ and of $\sqrt{s_{e^+ e^-}}$, and perform the {\it Signal/Background}
(S/B) analysis for the search for ``50 GeV neutrino" at LEP II and future
linear collider TESLA \cite{ee500}. The calculation has been performed
using the computer code CompHEP \cite{CompHEP}. 

The fourth generation leptons were introduced in the Standard Model
Lagrangian as an isodoublet for left-handed components of massive fermions
(including neutrino) and isosinglet for right-handed states. Then, we
assume vanishing mixing with the fourth generation -- mixing angle between
the fourth generation neutrino and those of first three generations should
be less than $10^{-6}$ \cite{100}.

Higher order QED corrections connected with radiation from initial electron
and positron states were taken into account by using the ``structure
function" technique: the squared matrix elements in the Born approximation
were convoluted with the {\it Initial State Radiation} (ISR) structure
function \cite{ISR} at the scale parameter $Q=\sqrt{s}$. We found that ISR
enlarges the signal cross sections by about 8-3\% for $M_N=46-50$ GeV,
while background is not changed or slightly decreased for these values of
neutrino mass because of opposite effect in $\nu_e\bar\nu_e$ and two other
(muonic an tauonic) channels. Note, that on the $Z$-peak shoulder this
enhancement is much higher, about 20-30\% both for the signal and
background. 

SM parameters were set in calculations as $M_Z=91.1884$ GeV,
$\Gamma_Z=2.45$ GeV, $M_W=80.343$ GeV. Couplings at electroweak vertices
were normalized to $\alpha(m_e)\cdot\alpha(M_Z)^2= (1/137)\cdot(1/128)^2$,
thus taking $W$ and $Z$ couplings at the electroweak scale while the photon
emission vertex is kept in the classical limit.

\vspace{0.5cm}
{\large\bf 2. \hspace{0.5cm} 50 GeV neutrinos at LEP II}
\vspace{0.3cm}

As we have mentioned already, the single photon final states were analyzed
by all LEP II experiments \cite{ALEPH,DELPHI,L3,OPAL} and no deviations
from SM predictions were observed. There were established limits on
additional photon production in the context of different ``new physics"
hypotheses, in particular in SUSY models and low scale gravity models with
extra dimensions. It is clear, that these data can be used also to
establish the limits on the mass of the fourth generation (Dirac) neutrino
in the discussed mass range $m_N\sim 50$ GeV, what was not done yet as we
know. In this section we make a phenomenological analysis of possible
limits on $m_N$ taking into account LEP II data, being collected already
\cite{ALEPH,DELPHI,L3,OPAL},

\begin{center}
\begin{tabular}{llll}
1997-1998 & $\sqrt{s}=182,189$ GeV,  & $\sim 226$ pb$^{-1}$/Coll.,  &
     ${\cal L}_{tot}=904.2$ pb$^{-1}$; \\
1999 & $\sqrt{s}=192-202$ GeV,  & $\sim 230$ pb$^{-1}$/Coll.,  &
     ${\cal L}_{tot}=923$ pb$^{-1}$,
\end{tabular}
\end{center}

\noindent
and data from current (final) run with collision energies $199.9-208.7$ GeV
(average energy 205 GeV). From LEPII schedule one can expect that it will
be delivered about 200  pb$^{-1}$ during the 2000 run. Thus, four
experiments will accumulate in total 800 pb$^{-1}$ additional data.

\begin{figure}[hb]
\unitlength=1cm
\vspace*{1.5cm}
\begin{picture}(16,14.5)
\put(0,8){\epsfxsize=8cm \leavevmode \epsfbox{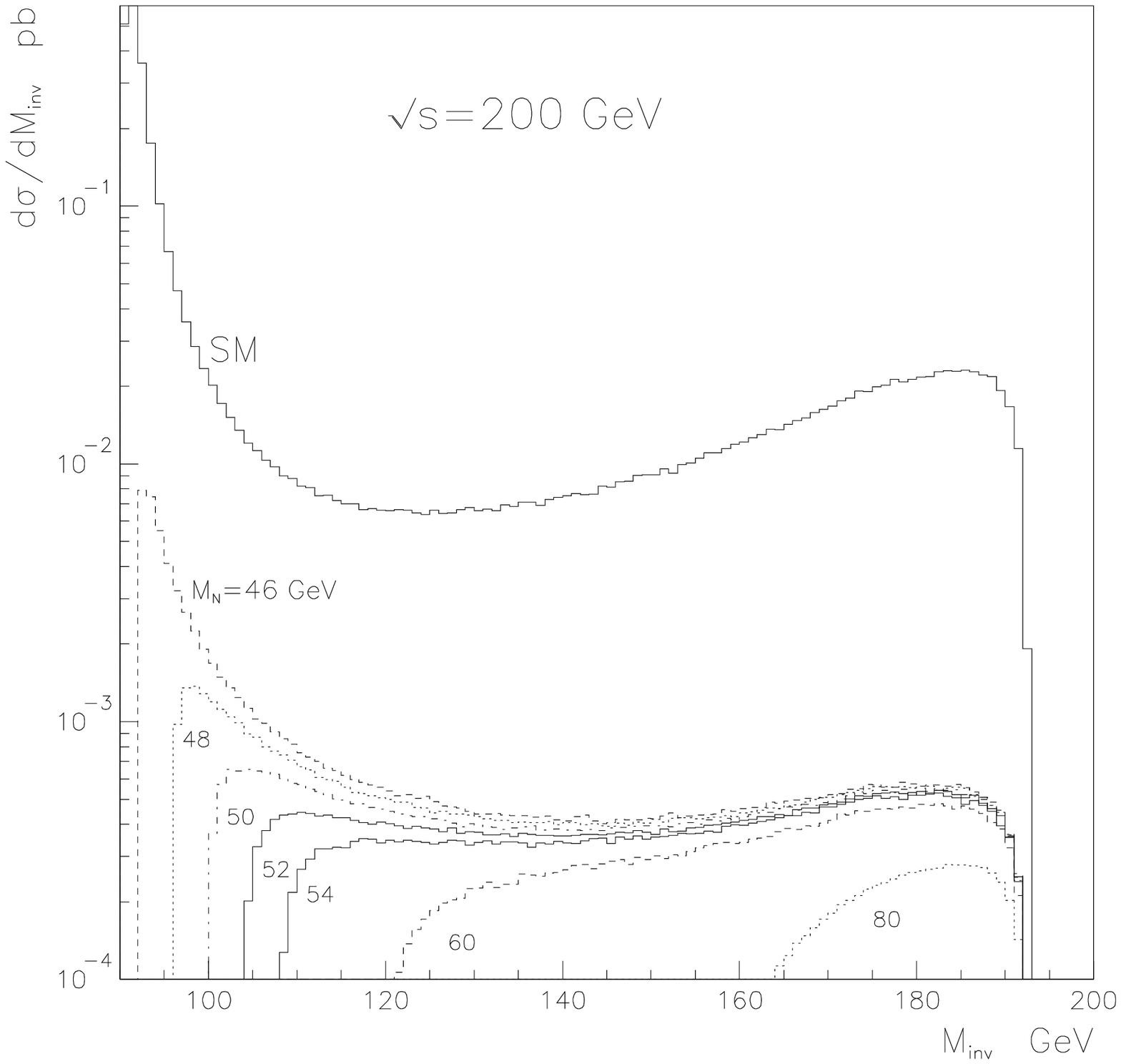}}
\put(8,8){\epsfxsize=8cm \leavevmode \epsfbox{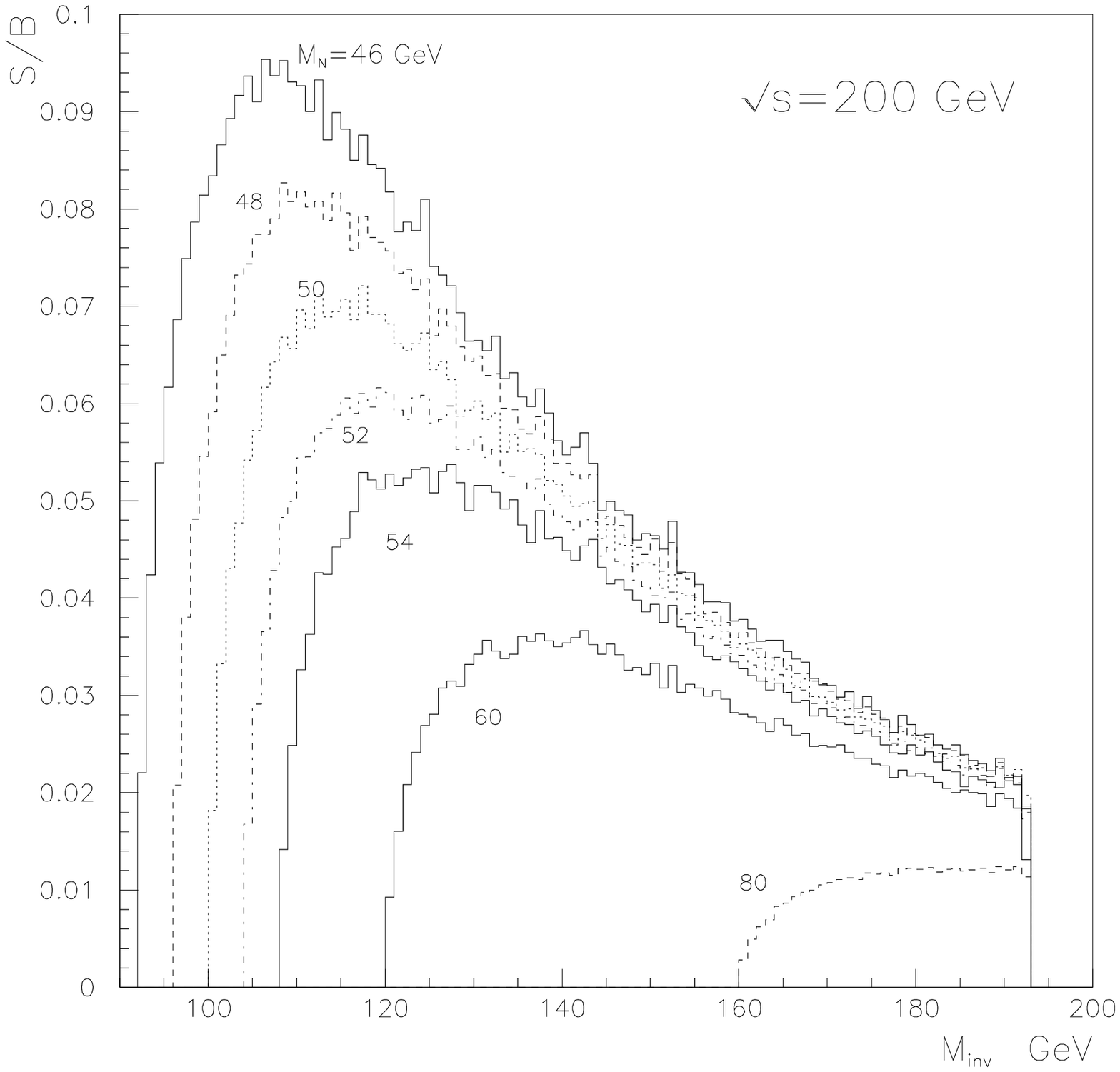}}
\put(0,0){\epsfxsize=8cm \leavevmode \epsfbox{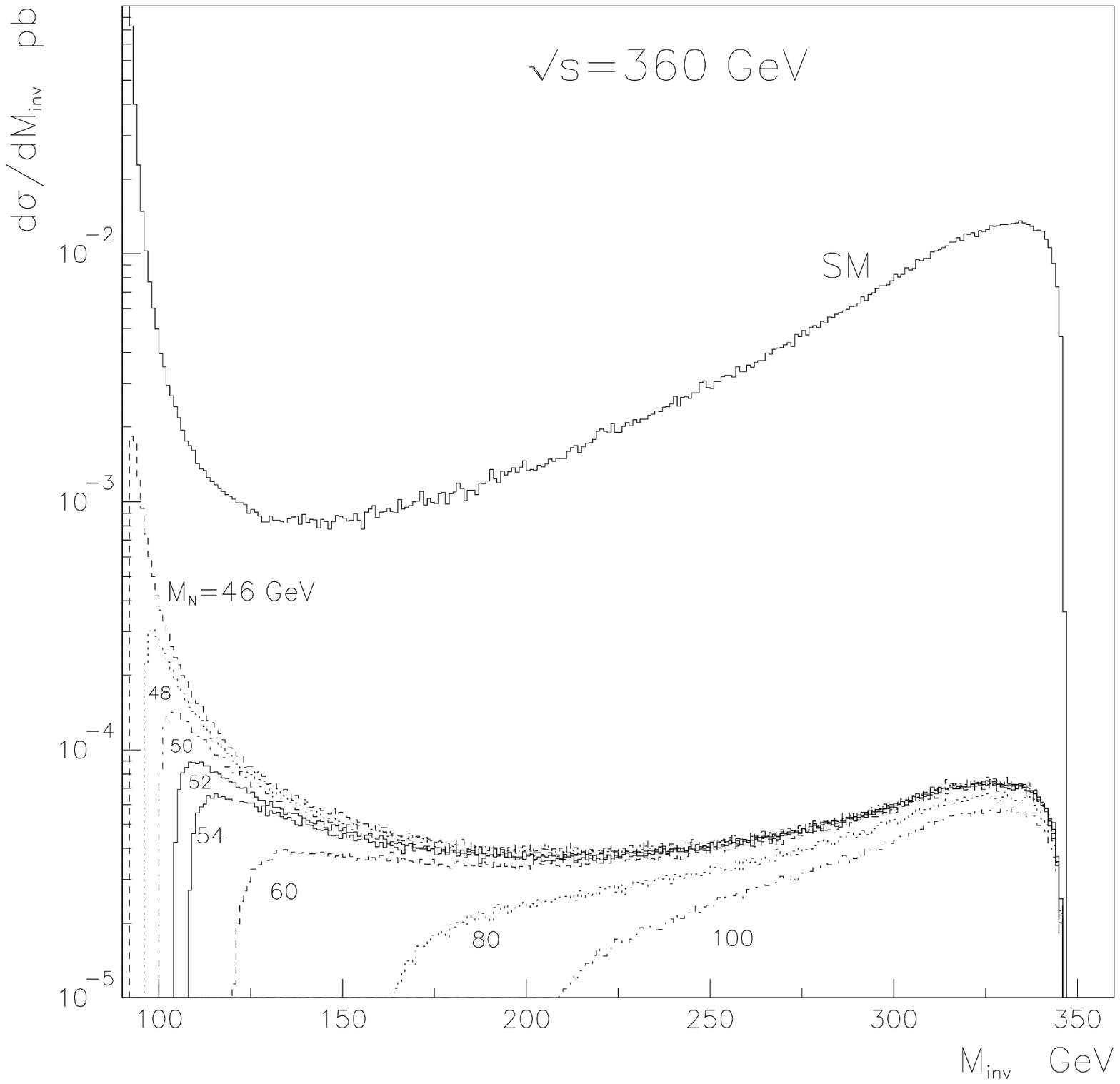}}
\put(8,0){\epsfxsize=8cm \leavevmode \epsfbox{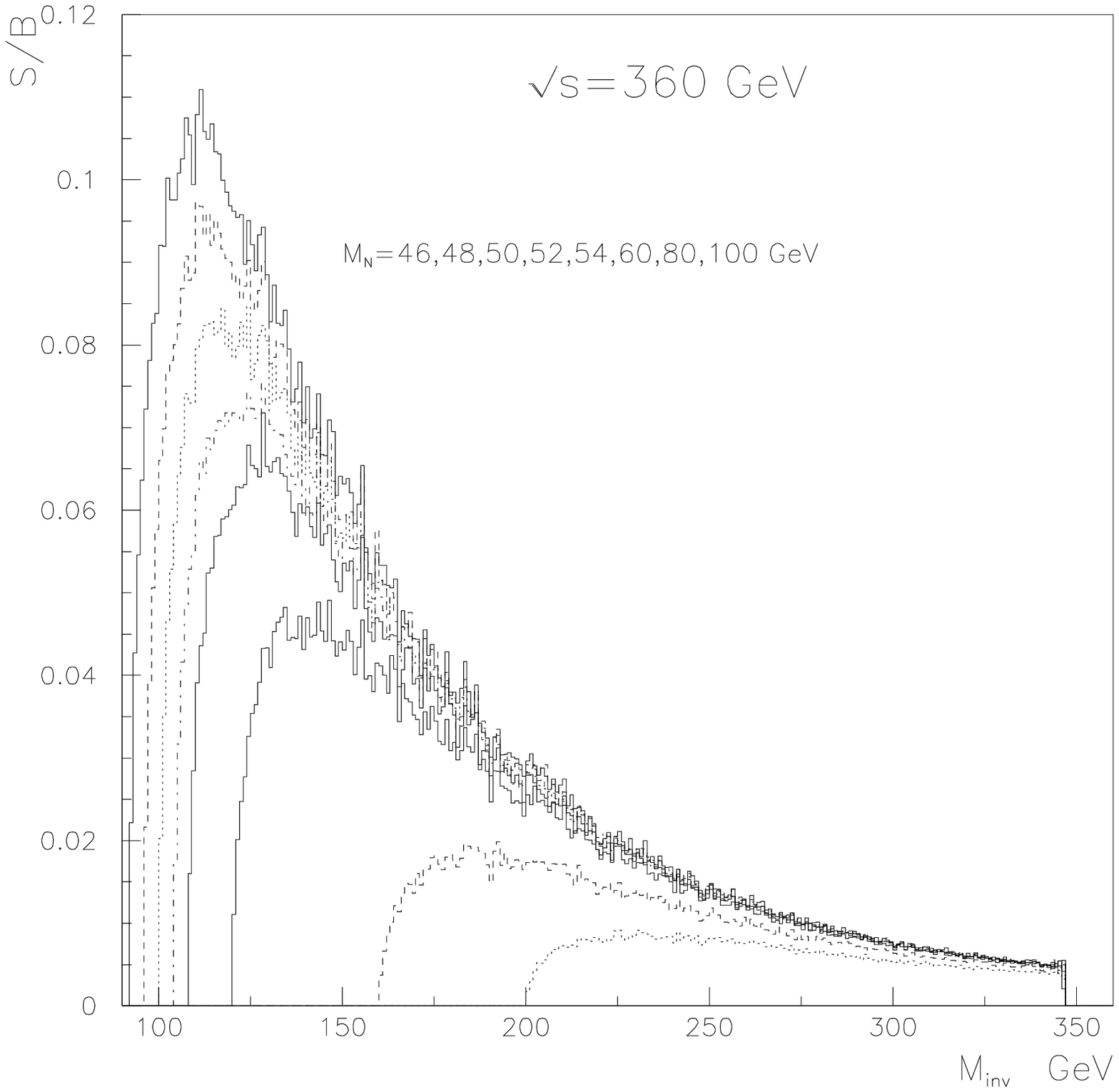}}
\end{picture}
\caption{
The left side plots: $d\sigma/d M_{inv}$ (in pb) for 
Standard Model and for the different values of $m_N$.
The right side plots: the {\it Signal/Background} ratio as a function of
$M_{inv}$.
Cuts applied: $|\cos\vartheta_\gamma|<0.95$ and $p_T^\gamma>0.0375\sqrt{s}$. 
The photon detection efficiency 74\% is assumed.
\label{fig:Minv200-360} }
\end{figure}

In Fig.~\ref{fig:Minv200-360} (upper left side plot) the distribution on
``invisible" mass $M_{inv}$ (invariant mass of the neutrino 
pair)\footnote{$M_{inv}$ is directly connected with another variable
frequently used, $x_\gamma\equiv E_\gamma/E_{beam}=1-M_{inv}^2/s$.} is
represented for SM background and the $N\bar N$ signal for $\sqrt{s}=200$
GeV and different values of $N$ masses, $M_N=46-100$ GeV. Here we applied
kinematical cuts on the photon polar angle and transverse momentum,
$|\cos\vartheta_\gamma|<0.95$ and $p_T^\gamma>0.0375\sqrt{s}$, being the
ALEPH selection criteria \cite{ALEPH}. Other experiments tried to include
in the analysis also events with smaller photon polar angles and lower
transverse momenta. E.g. DELPHI collected events with photon emmited at
angles up to $3.8^\circ$. 

One can see that the S/B ratio is better on the $Z$-peak shoulder. However,
for higher significance of the $N\bar N$ signal, evaluated as
$N_S/\sqrt{N_B}$, one should include whole interval on $M_{inv}$ allowed
kinematically. For example, for statistics 230 pb$^{-1}$ the signal
significance is 0.72 for $m_N=48$ GeV, see Table~\ref{tab:sign200} for
details. In this case SM predicts about 212 events per experiment with only
9 events with massive neutrino pair. Here we have taken into account the
photon detection efficiency -- in the ALEPH Monte-Carlo analysis
\cite{ALEPH} it was estimated at the level 74\%. For other experiments the
photon detection efficiency is lower, but they used wider selection
criteria.

Let us now look how the $N\bar N$ signal significance depends on angular
and transverse momentum cuts. In the Table~\ref{tab:cuts200} the cross
sections for SM background and $N\bar N$ signal (with $m_N=48$ GeV) are
given for three values of the angular and $p_T$ cuts. One can see that the
signal significance is higher for weaker cuts.

\begin{figure}[hb]
\unitlength=1cm
\vspace*{1.5cm}
\begin{center}
\begin{picture}(16,8)
\put(0,0){\epsfxsize=8cm \leavevmode \epsfbox{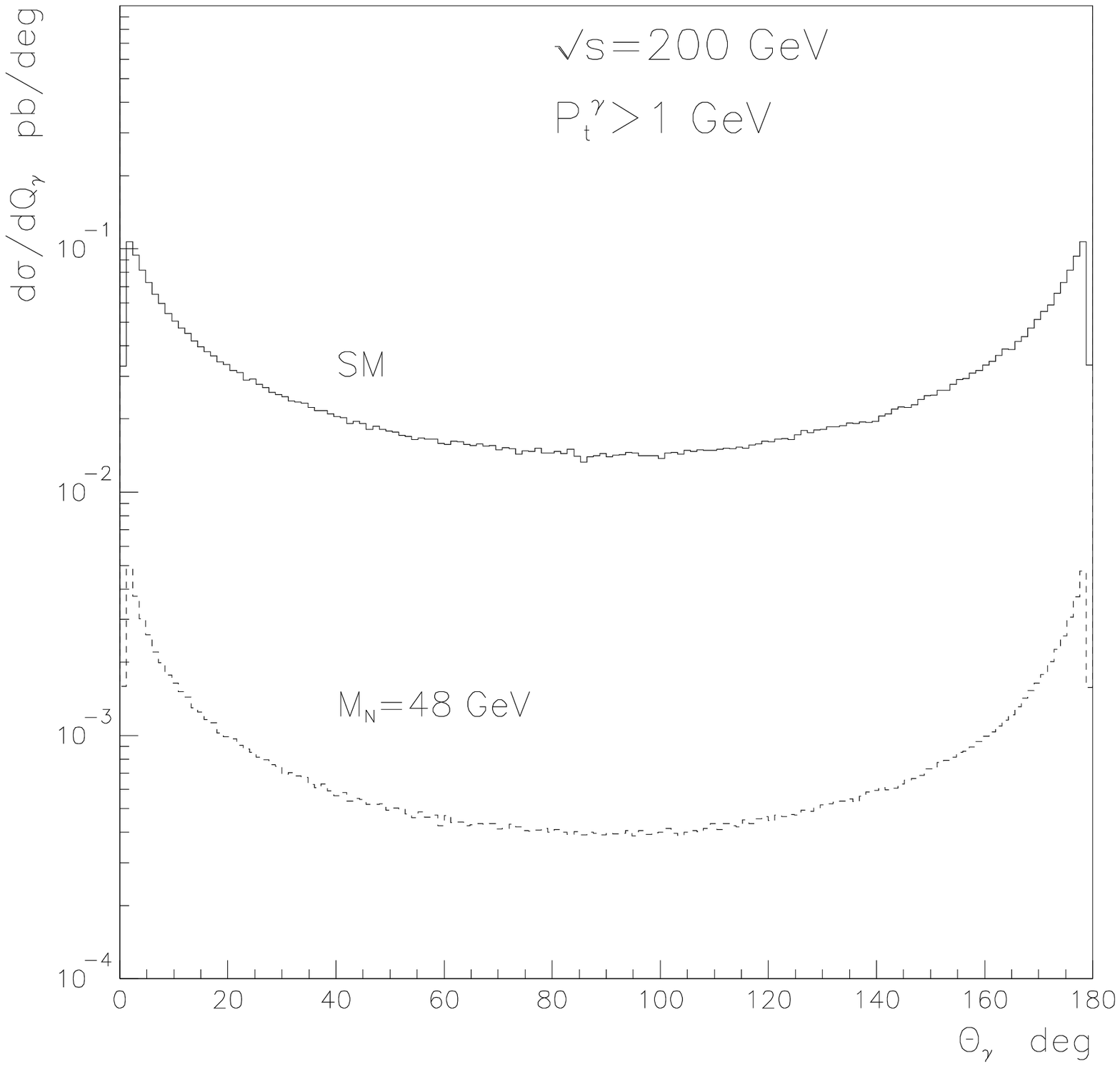}}
\put(8,0){\epsfxsize=8cm \leavevmode \epsfbox{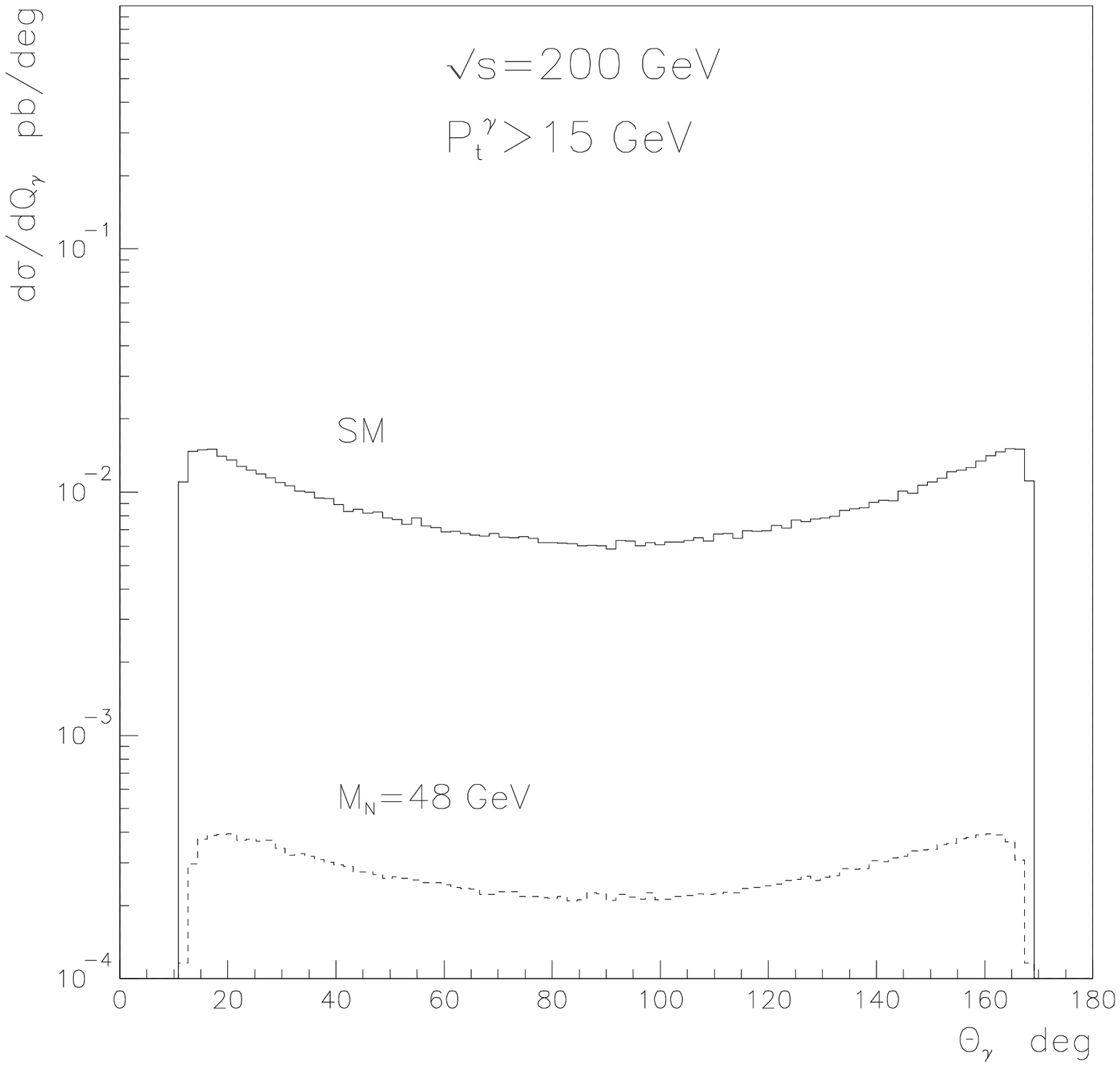}}
\end{picture}
\end{center}
\caption{Distribution on photon polar angle for the Standard Model and for
the $N\bar N$ signal with $m_N=48$ GeV. The cut $M_{inv}>2m_N$ is applied.
\label{fig:theta200} }
\end{figure}

Then, we found that, in spite of the $N\bar N$ signal being due to
$s$-channel diagrams, while large contribution to the SM background is due
to $t$-channel $W$-exchange, the angular distributions have similar shape
for signal and background; that is clearly seen on Fig.~\ref{fig:theta200}.
Thus, it seems that there are no chances to improve the signal significance
by using more complicated selection criteria, e.g. exploiting the
correlation between photon polar angle and $p_T$.

The general conclusion from this analysis is that one should collect single
photon events with the weakest kinematical cuts allowed by detector
construction and conditions to suppress reducible background (e.g. Bhabha
scattering). Of course, only a careful simulation can give answer what are
the optimal selection criteria for the search for ``50 GeV neutrino". In
what follows we use ALEPH selection criteria, $|\cos\vartheta_\gamma|<0.95$
and $p_T^\gamma>0.0375\sqrt{s}$, and apply cut $M_{inv}>2m_N$.

\begin{figure}[hb]
\unitlength=1cm
\begin{center}
\begin{picture}(16,16)
\put(0,8){\epsfxsize=8cm \leavevmode \epsfbox{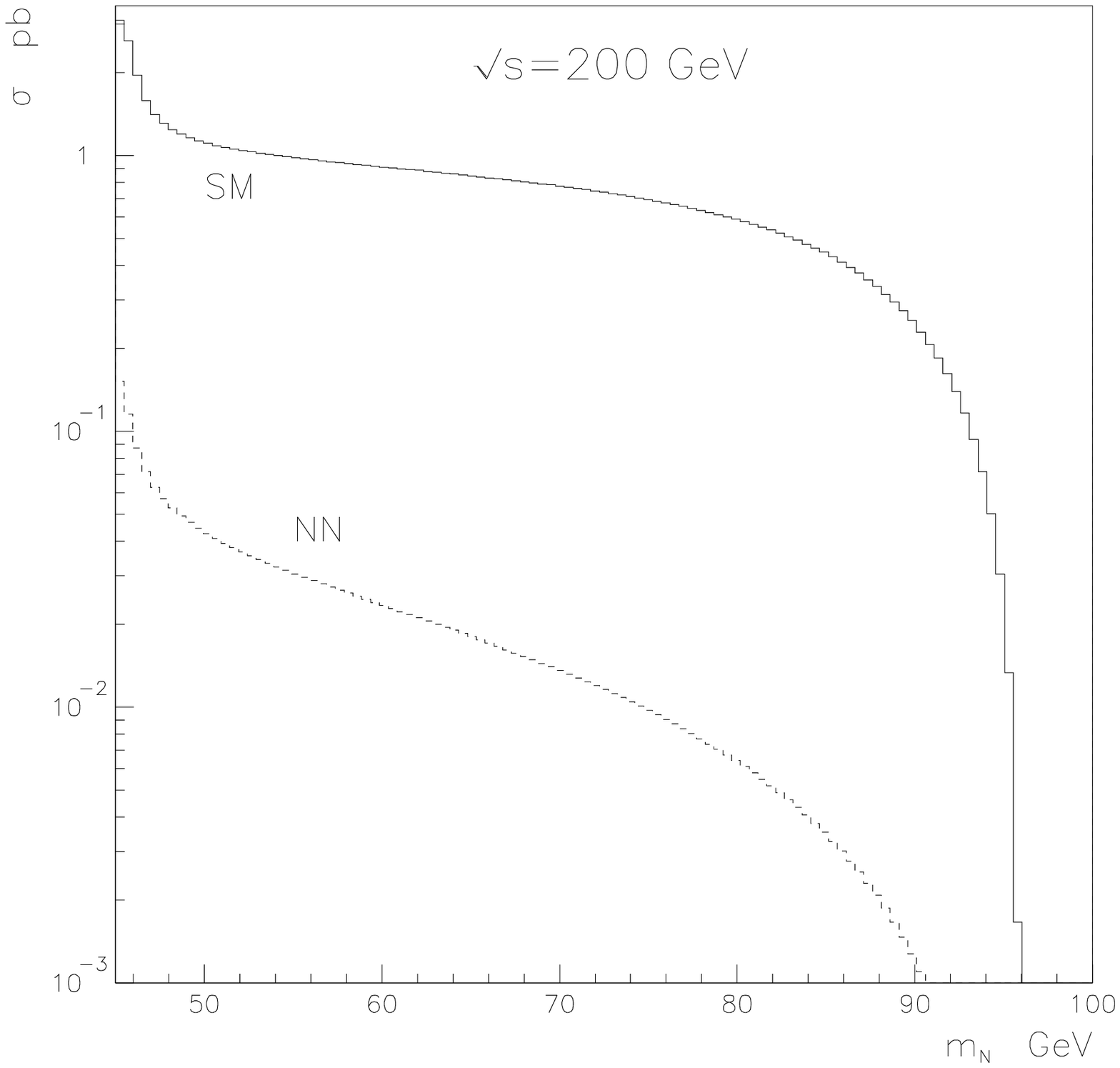}}
\put(8,8){\epsfxsize=8cm \leavevmode \epsfbox{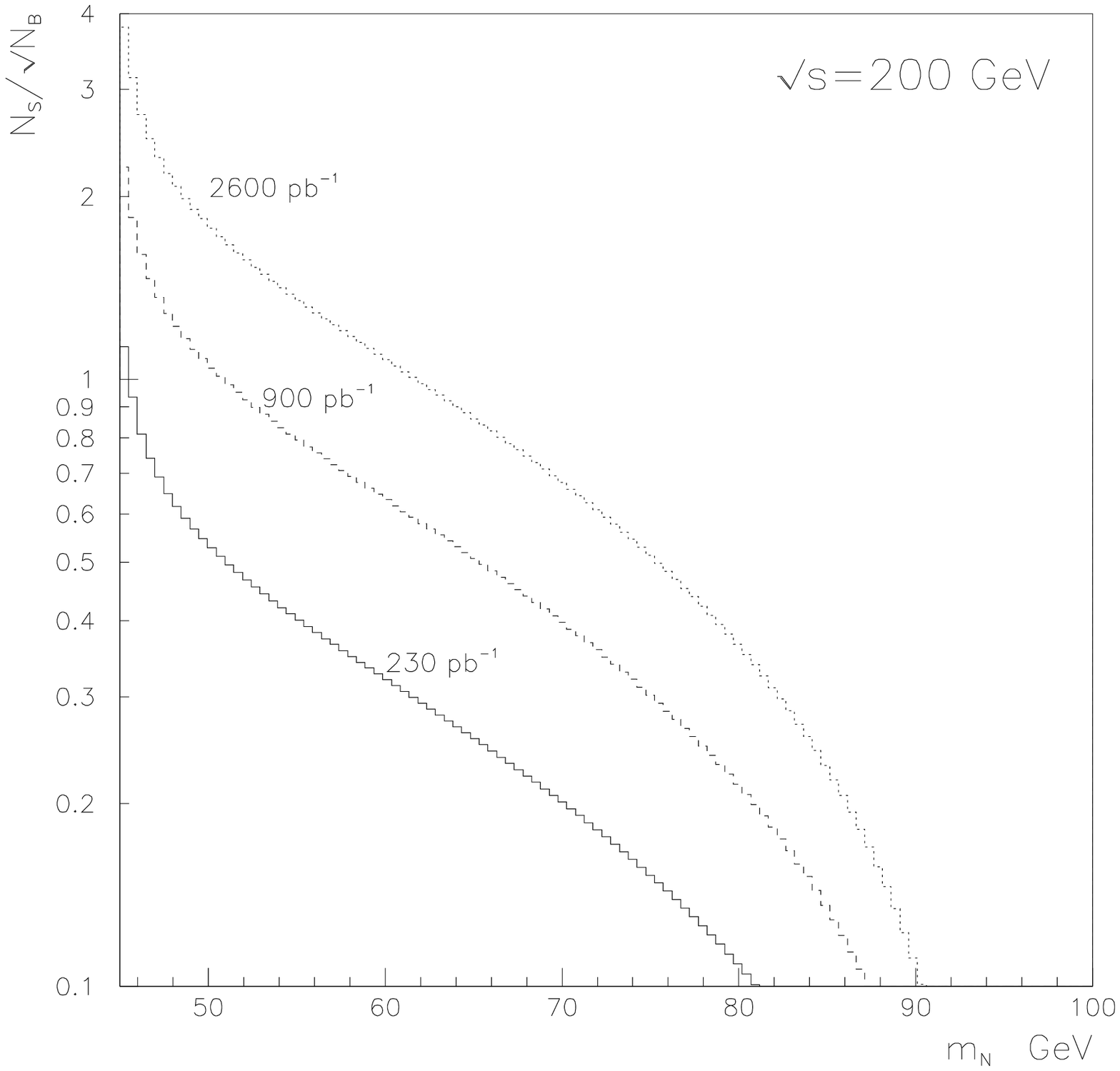}}
\put(0,0){\epsfxsize=8cm \leavevmode \epsfbox{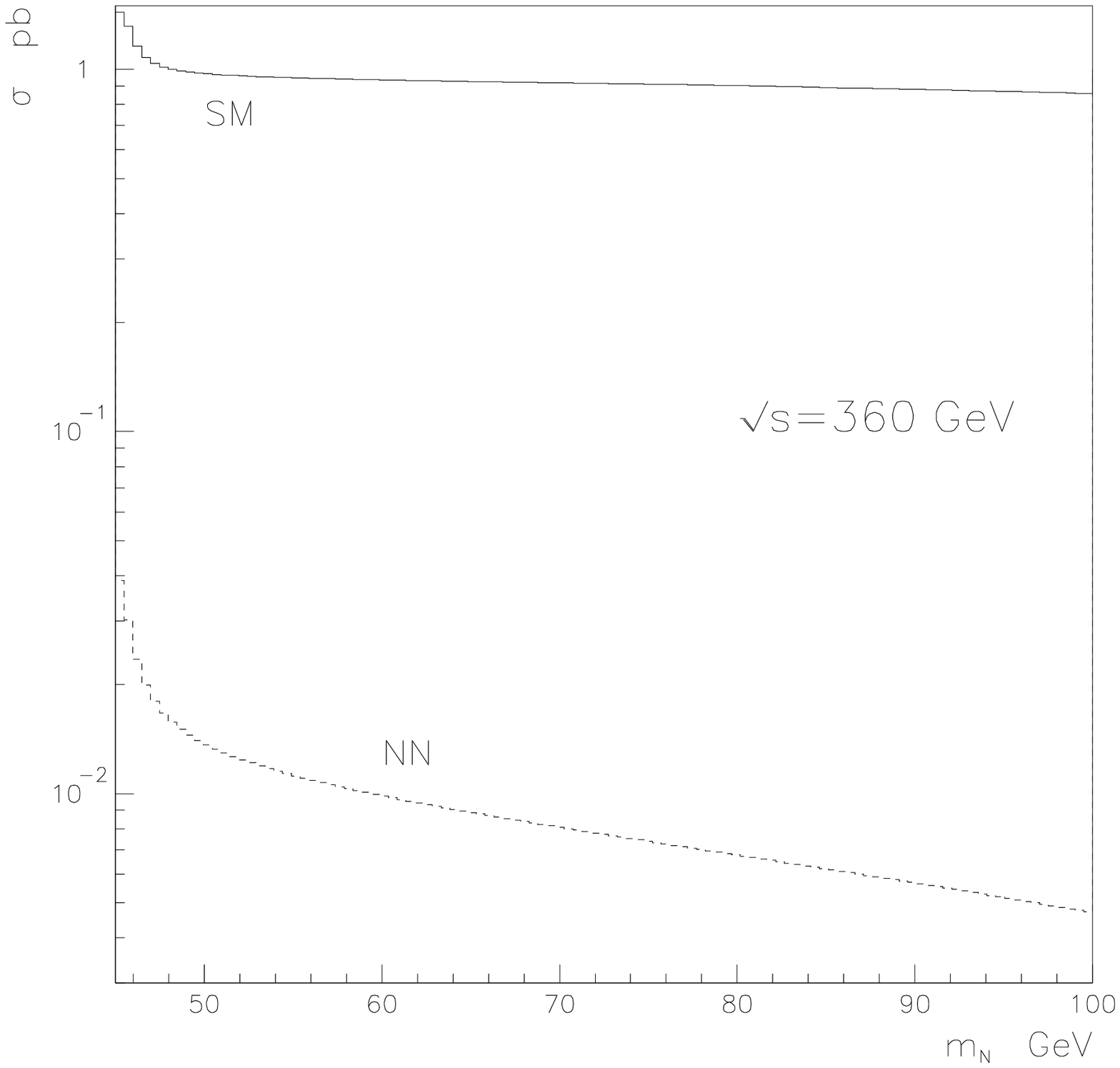}}
\put(8,0){\epsfxsize=8cm \leavevmode \epsfbox{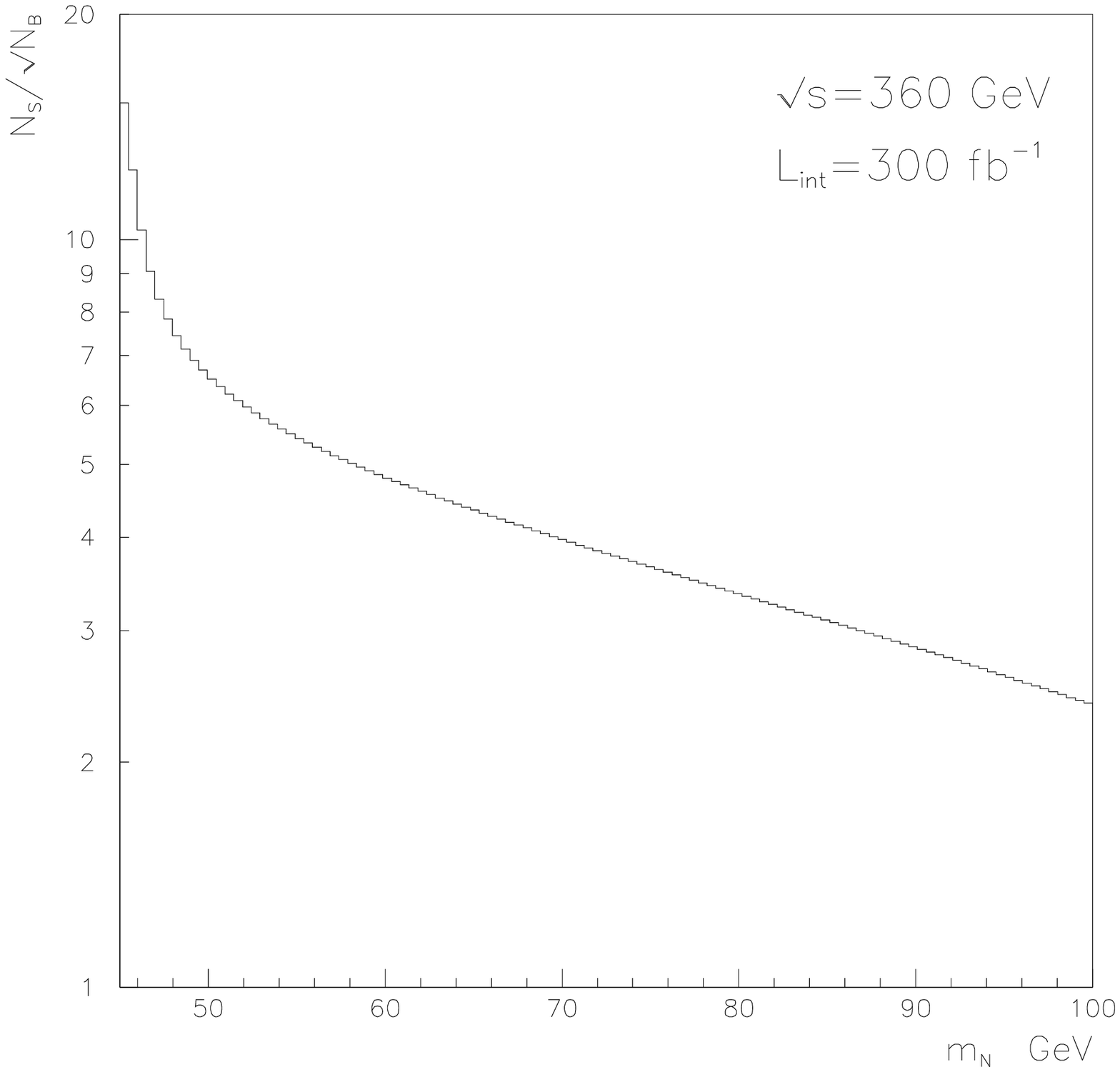}}
\end{picture}
\end{center}
\caption{Cross sections for the $N\bar N$ signal and SM background (left side plots)
and $N\bar N$ signal significances at different statistics (right side plots)
as function of the neutrino mass. 
Cuts applied: $M_{inv}>2m_N$, $|\cos\vartheta_\gamma|<0.95$ and
$p_T^\gamma>0.0375\sqrt{s}$.
For the significance the photon detection efficiency 74\% is assumed.
\label{fig:cs-mN} }
\end{figure}

On Fig.~\ref{fig:cs-mN} the signal and background cross sections and signal
significances (for different samples of data) are represented as a function
of $m_N$. One can derive that only the analysis based on combined data from
all four experiments both from 1997-1999 runs ($\sqrt{s}=182-202$ GeV) and
from the current run, in total $\sim 2600$ pb$^{-1}$, can exclude at 95\%
CL the interval of $N$ mass up to $\sim 50$ GeV.

\vspace{0.5cm}
{\large \bf 3. \hspace{0.5cm} 50 GeV neutrinos at TESLA}
\vspace{0.3cm}

 From the point of view of expected cross sections the increase in energy
leads to the decrease both of the signal and the background. This can be
seen from comparing the left-side plots on Fig.~\ref{fig:cs-mN}. However
this decrease may be over compensated by the proposed increase of
luminosity of the future linear $e^+ e^-$-collider. Thus for TESLA the
expected year luminosity at the first stage ($\sqrt{s}=360$ or 500 GeV) is
300 ${\rm fb}^{-1}$ \cite{TESLA}.

Further advantage of the linear collider is the possibility to use
polarized beams (80\% circular polarization for electrons and 60\% for
positrons). This is important in suppressing the cross section of $e^+ e^-
\to \nu_e\bar{\nu}_e\gamma$ as this reaction goes mainly through the
$t$-channel exchange of the $W$ boson and gives substantial contribution to
the background\footnote{For ``light" neutrinos $45 \; {\rm GeV} < M_N < 50$
GeV the ratio of $s$ channel to $t$ channel contributions is enhanced at
the shoulder of $Z$-peak.}. However, even without exploiting the beam
polarization the advantage of TESLA in the total number of events is
extremely important. Thus, Standard Model is expected to give approximately
0.3 million single photon events for $M_{inv}>100$ GeV while the number of
50 GeV neutrino pairs would be about 4000.  On Fig.~\ref{fig:cs-mN} (lower
plots) the signal and background cross sections, as well the significance,
are represented for TESLA as a function of $N$ mass. Here we applied cuts
$|\cos\vartheta_\gamma|<0.95$, $p_T^\gamma>0.0375\sqrt{s}$ and
$M_{inv}>2m_N$, and assumed photon detection efficiency to be 74\%,
similarly to the LEP II case.

With such numbers, although the signal over background ratio is still small
(2.3-0.5\% for $m_N=45-100$ GeV correspondingly) the significance of the
signal is excellent, higher than 2 for the whole interval of N neutrino
mass under discussion. It means that fourth generation Dirac neutrino will
be excluded at the 95\% CL in TESLA experiment. However, if such a neutrino
exists this new particle will be discovered with $5\sigma$ significance
after one year of the TESLA operation at first stage if $m_N<60$ GeV.

\vspace{0.5cm}
{\large\bf Conclusions}
\vspace{0.3cm}

 The total sample of single photon events from all four LEP II detectors
can exclude the 50 GeV neutrinos at the 95\% CL or show the evidences of
their existence with only $2\sigma$ significance.

The future TESLA experiment at its first stage (with collision energy 360
or 500 GeV) will be able to exclude or discover fourth generation, if one
combines TESLA results with the bounds from radiative corrections \cite{2}.

\vspace{0.5cm}
{\large\bf Acknowledgements}
\vspace{0.3cm}

We thank P.Coyle, A.Tilquin, A.Ealet and R.Tanaka for useful discussions.

The work of V.N., L.O. and M.I. have been supported by RFBR grants
98-02-17372 and  00-15-96562, V.N. and M.V. are supported by RFBR grant
98-02-17453 as well. L.O. acknowledges the A. von Humboldt Award. The work
of M.M. is supported by DGICYT under grant PB98-0693 and by the TMR network
grant ERBFMRX-CT96-0090 of the European Union. The work of V.I. has been
partially supported by the CERN-INTAS 99-377 and RFBR-DFG 99-02-04011
grants and by St.Petersburg Grant Center.


\clearpage
{\large\bf Tables}

\vspace{2cm}

\begin{table}[hb]
\begin{center}
\begin{tabular}{|l|llllll|}
\hline
max $M_{inv}$ GeV &  - & - & 140 & 130 & 120 & 110 \\ 
min $M_{inv}$ GeV & $2 m_N$ & 97.5 & 97.5 & 97.5 & 97.5 & 97.5 \\
\hline
$\sigma_{SM}$ pb    & 1.243 & 1.175 & 0.402 & 0.333 & 0.268 & 0.196 \\
$\sigma_{N\bar N}$ pb    &0.0529 &0.0512 &0.0269 &0.0228 &0.0182 &0.0125 \\
\hline
$N_S/\sqrt{N_B}$, 230 pb$^{-1}$ &0.720 &0.716 &0.644 &0.598 &0.571 &0.428 \\
\hline
\end{tabular}
\end{center}
\caption{Cross sections and $N\bar N$ signal significance for
different values of the $M_{inv}$ cut. 
$\sqrt{s}=200$ GeV and $m_N=48$ GeV.
Cuts applied: $|\cos\vartheta_\gamma|<0.95$ and
$p_T^\gamma>0.0375\sqrt{s}$.
\label{tab:sign200}}
\end{table}

\vspace{2cm}

\begin{table}[hb]
\vspace{2cm}
\begin{tabular}{|l|lll|lll|lll|}
\hline
$\vartheta_\gamma>$      & \multicolumn{3}{c|}{3} 
                      & \multicolumn{3}{c|}{18.2}
                      & \multicolumn{3}{c|}{30} \\
          deg         &  \multicolumn{3}{c|}{}
       &  \multicolumn{3}{c|}{($|\cos\vartheta_\gamma|<0.95$)}
                      &  \multicolumn{3}{c|}{} \\
\hline
$p_T^\gamma>$    & 1 & 7.5 & 15 & 1 & 7.5 & 15 & 1 & 7.5 & 15 \\
   GeV                & &  &  &  &  &  &  &  &  \\
\hline
$\sigma_{SM}$pb&6.089 &2.731 &1.682 &3.620 &1.947 &1.410 &2.668 &1.458 &1.082\\
$\sigma_{N\bar N}$pb&0.153 &0.0706&0.0436&0.0903&0.0527&0.0392&0.0668&0.0400
              &0.0304\\
\hline
$N_S/\sqrt{N_B}$,  
               & 0.942 & 0.648 & 0.509 & 0.720 & 0.573 & 0.501 & 0.620 &
               0.502 & 0.443\\
230 pb$^{-1}$  & & & & & & & & & \\
\hline
\end{tabular}
\caption{Cross sections and $N\bar N$ signal significance
for different values of the cuts on photon polar angle and transverse
momentum.
$\sqrt{s}=200$ GeV, $m_N=48$ GeV.
The cut $M_{inv}>2m_N$ is applied.
\label{tab:cuts200}}
\end{table}

\clearpage
{\large\bf Figures}

\end{document}